\documentstyle[PASJadd]{PASJ95}
\begin{document}

\title{Molecular Clouds Around a Run-away O 
Star, $\zeta$ Oph}

\author{ Kengo {\sc Tachihara},$^{1,2}$
Rihei {\sc Abe},$^2$ 
Toshikazu {\sc Onishi},$^2$ 
Akira {\sc Mizuno},$^2$ 
and Yasuo {\sc Fukui}$^2$ 
\\[12pt]
$^1${\it Max-Planck-Institut f\"ur extraterrestrische Physik, 
Giessenbachstra\ss e, D-85748 Garching, Germany}\\
{\it  E-mail(KT): tatihara@mpe.mpg.de}\\
$^2${\it Department of Astrophysics, Nagoya University, 
Chikusa-ku, Nagoya 464-8602, Japan}}

\abst{
Molecular clouds around a run-away O star $\zeta$ Oph have been 
surveyed with NANTEN telescope and their streaming motion 
caused by $\zeta$ Oph has been detected.  $\zeta$ Oph is the earliest 
(O9.5V) member of the Sco OB2 association and is a runaway star 
rapidly moving accompanied by an H\,{\footnotesize II} region S27.  
We detected 2 major filamentary cloud complexes; one complex 
including L156 (L156 complex) is lying across nearly the center 
of S27 and the other one (L204 complex) is located near the 
eastern edge of S27.  Total masses of them traced by the $^{12}$CO 
emission in the two complexes are 520 $M_{\odot}$ and 1110 
$M_{\odot}$, respectively.  
Denser molecular cloud cores detected in C$^{18}$O are locally 
distributed on the near side of the L204 complex to $\zeta$ Oph, 
and lower density gas traced by $^{12}$CO spreads toward the 
opposite side.  Both complexes have radial velocity shifts that 
are correlated with the gas density.  These spatial and 
velocity structures can be interpreted as follows; (1) the L156 
complex is stuck on the expanding Str\"omgrem sphere and 
has been accelerated, (2) the molecular gas in L204 complex was 
compressed and has also been accelerated outward from the 
H\,{\footnotesize II} region by $\zeta$ Oph, resulting the radial 
velocity shifts of diffuse low-density gas relative to the dense 
cores embedded in the cloud.  

These density and velocity structures indicate dynamical interaction 
between the H\,{\footnotesize II} region and the molecular clouds.  
The cloud complexes are divided into seven clouds by intensity 
distributions.  In order to investigate the acceleration mechanism, 
we calculated momentum and kinetic energy for each cloud.  They 
range from 60 to 800 $M_{\odot}$ km s$^{-1}$ and from 0.9 to 
21 $\times 10^{45}$ erg, respectively.  We examined the effects of 
the stellar wind and photo evaporation by UV field of $\zeta$ Oph 
and found that the stellar wind can hardly input the momentum 
during the crossing time of the rapid movement of $\zeta$ Oph.  UV 
radiation seems to be a more likely origin of the streaming gas 
motion.
}

\kword{Interstellar: clouds --- Interstellar: individual 
(Ophiuchus region, zeta Oph, L156, L204) --- 
Interstellar: kinematics and dynamics --- Interstellar: molecule 
--- radio lines: Interstellar --- stars: formation}

\maketitle
\thispagestyle{headings}

\section{Introduction}

$\zeta$ Oph (HD149757; O9.5V type, $\alpha_{1950} = 16^{\rm h} 
34^{\rm m} 24^{\rm s}\hspace{-5pt}.\hspace{2pt}1$, $\delta_{1950} 
= -10^{\circ} 28' 03''$) is the earliest star among the members of 
Sco OB2 association (de Geus et al.\ 1989).  
The bright UV and optical light of the star provide good 
opportunities to study foreground diffuse interstellar medium in 
many emission and absorption lines of various molecules and 
atoms, and chemical composition has been derived (e.g., Morton 
1975, Langer et al.\ 1987, Kopp et al.\ 1996, and Liszt 1997).  
$\zeta$ Oph itself is ionizing a density-bound H\,{\footnotesize II} 
region, S27, which spreads in an elliptical shape of $\sim 7^{\circ} 
\times 10^{\circ}$ in $\alpha$ and $\delta$ (Morgan et al.\ 1955), 
corresponding to $\sim$ 18 pc $\times$ 26 pc at a distance of 140 pc.  
Some of the extinction features shading the H$\alpha$ emission are 
seen in a plate taken by Sivan (1974) which represents the existence 
of dark clouds in the foreground of S27.  These features coincide with 
the dark clouds L156, L204, L190, and so on (Lynds 1962), and clouds 
may be interacting with the H\,{\footnotesize II} region.  

Physical interactions between early type stars and molecular 
clouds should be important processes related to formation, 
evolution, and dissociation of molecular clouds.  
$\zeta$ Oph and its surroundings are one of the best sites to 
investigate these interactions because of its proximity to the sun.  
The distance to the star is estimated to be 200 pc (Lesh 1968), 170 pc 
(Bohlin 1975), and 140 pc (Draine 1986).  
$\zeta$ Oph is also known as a rapidly moving run-away star, and 
the proper motion is measured as $\mu_{\alpha*} = 0.013''$ yr$^{-1}$, 
$\mu_{\delta} = 0.025''$ yr$^{-1}$ (Perryman et al.\ 1997).  
The helio-centric radial velocity of $\zeta$ Oph is $-10.7$ km 
s$^{-1}$ (Lesh 1968) and a space velocity with respect to the local 
standard of rest (LSR) is 3.3 km s$^{-1}$ in the radial direction, 
31.2 km s$^{-1}$ in the direction of increasing $l$, and 5.0 km 
s$^{-1}$ in the direction of increasing $b$ (Draine 1986).  $\zeta$ 
Oph, therefore, has passed through the Ophiuchus region for a few 
millions years.  This implies that the molecular clouds may have 
experienced being illuminated by strong UV light in a short time 
scale, and this gives us an ideal laboratory to study the physical 
interaction between molecular clouds and an early type star.

McCutcheon et al.\ (1986) made mm-wave 
observations in the $J=$ 1--0 $^{12}$CO and $^{13}$CO emission 
toward L204, and found that the cloud has velocity structures 
which correlate with spatial distribution of the bent filament.  
They suggested that the filamentary cloud is influenced by 
external compression, possibly due to $\zeta$ Oph, resulting 
the velocity structure and the morphology of the cloud.  
The spatial and density coverage is, however, limited and 
more extensive observations with wide density regime are 
needed to reveal the dynamical interaction.  
They also showed that the optical polarization vectors are 
aligned perpendicular to the long axis of L204, which suggests 
the magnetic fields penetrating the filamentary cloud 
perpendicularly.  Heiles (1988) estimated the magnetic field 
strength from the H\,{\footnotesize I} Zeeman splitting to be 
$\sim 12 \mu$G.  

Nozawa et al.\ (1991) made extensive $^{13}$CO observations 
covering the entire Ophiuchus North region and revieled large 
scale cloud distribution.  There are 3 molecular clouds identified 
in an area of $\sim$ 10 deg$^2$ around $\zeta$ Oph, while only one 
young stellar object is found to be physically associated there.  
They conclude that the Ophiuchus North region is inactive in star 
formation despite of the existence of massive molecular gas of 
$\sim 4400 M_{\odot}$.  Subsequently, denser gas distributions 
were studied by Tachihara et al.\ (2000) with C$^{18}$O observations, 
that found the dense cloud cores whose average density is $\sim10^4$ 
cm$^{-3}$ embedded in the molecular clouds.  

In order to reveal the dynamical interaction in the region of $\zeta$ 
Oph and to investigate how it affects the cloud structure and 
dynamics, we have made extensive $^{12}$CO molecular line 
observations toward the region.  In section 2, we briefly summarize 
the observational properties and spatial and velocity distributions, 
and the physical parameters of the detected clouds are mentioned 
in section 3.  Section 4 discusses the dynamical interaction 
between $\zeta$ Oph and the molecular clouds by introducing the 
calculations of kinematic energy and momentum.  Finally, the 
summary of this paper is given in section 5.

\section{Observations}

An area of $\sim 47$ deg$^2$ around $\zeta$ Oph and S27 has been 
observed in $^{12}$CO ($J = $1--0) emission line (115.2 GHz) with the 
NANTEN millimeter wave telescope (HPBW = $2'\hspace{-4.5pt}.\hspace
{.5pt}7$) at Las Campanas observatory.  In total, 10575 $^{12}$CO 
spectra were obtained with $4'$ grid spacing.  Velocity resolution of 
each spectrum is $\sim$ 0.1 km s$^{-1}$, and velocity coverage is 
100 km s$^{-1}$ centered at 5 km s$^{-1}$.  Absolute antenna 
temperature was calibrated by adopting the peak $T_{\rm r}^*$ of 
$\zeta$ Oph East, $IRAS$ point source 16293$-$2422, as 15 K.  The 
typical integration time for each observed point was $\sim$ 5 sec, 
and rms noise temperature after the calibration is $\sim$ 0.56 K for 
0.1 km s$^{-1}$ velocity resolution.  

To investigate the denser region of the molecular cloud, L204 
was observed also in $^{13}$CO ($J = $1--0) emission line (110.20137 
GHz) with the 4m millimeter wave telescope installed in Nagoya 
University, and 491 $^{13}$CO spectra were obtained with $4'$ grid 
spacing.  Velocity resolution is the same as above.  The intensity scale 
of $^{13}$CO spectra were calibrated referring $T_{\rm r}^*$ of 
M17SW as 14.7 K (Nozawa et al.\ 1991).  In order to compare dense 
cores embedded in the cloud, C$^{18}$O data were also obtained with 
the same telescope for the intense parts of the $^{13}$CO emission 
with $2'$ grid spacing (Tachihara et al.\ 2000).  The total number of 
C$^{18}$O spectra obtained in this region is $\sim$ 940.  

\section{Results}

\subsection{Overview of the $^{12}$CO distribution}

The integrated intensity of the $^{12}$CO ($J = $1--0) emission is 
shown in Fig. 1.  Two major filamentary complexes exist lying from 
the north to the south.  The western one runs along $\alpha = 
16^{\rm h} 33^{\rm m}$ crossing through the center of the 
H\,{\footnotesize II} region with $\sim$ 10-pc length and $\sim$ 
3-pc width.  Some dark clouds including L156, L145, and L121 
were identified inside of this complex (hereafter L156 complex).  
$\zeta$ Oph exists at $\sim$ 1 degree north of the center of L156 
complex and the $^{12}$CO intensity toward the star is apparently 
weaker than the other parts.  Another cloud complex (hereafter L204 
complex) has a curved filamentary structure of $\sim$ 20-pc 
length and $\sim$ 2-pc width as a whole, being located near the 
eastern boundary of the H\,{\footnotesize II} region.  Some $^{12}$CO 
peaks are also embedded in places, being connected by filamentary 
clouds.  Diffuse components spread toward the east of the L204 
complex.  Little CO emission is, however, detected between the 2 
complexes.  At the western edge of the L204 complex, a large intensity 
gradient is shown by tight contour lines, suggesting a relatively large 
density gradient there.  The $^{12}$CO integrated intensity is stronger 
in the L204 complex than in the L156 complex on the whole; 
the strong intensity at the clumpy regions in the L204 complex is 
especially prominent.

\subsection{Physical parameters}

L204 complex is divided into 6 clouds bordered by less intense 
bridges (see Fig. 2).  Regarding L156 complex as one cloud, 
we estimate physical parameters for each cloud as show in Table 
1.  In general, $^{12}$CO ($J = $1--0) emission is optically thick.  It 
becomes saturated when the H$_2$ column density, $N$(H$_2$), is 
larger than $10^{21}$ cm$^{-2}$, and may not reflect the actual 
$N$(H$_2$).  Nonetheless, we estimate physical parameters of 
$N$(H$_2$), mass, and density using an empirical relation, $N({\rm 
H_{2}}) = W({\rm ^{12}CO}) \times 1.56 \times 10^{20}$ cm$^{-2}$ 
(Hunter et al.\ 1997) where $W$($^{12}$CO) is the integrated 
intensity of CO emission.  Note that this conversion factor may 
change from region to region according to the circumstances.  
In the region with strong UV field, the value is expected to be 
larger because of higher degree of CO dissociation.  It is, however, 
very difficult to estimate the factor for each cloud, so here we 
assume that it is uniform all through the region.  In this sense, 
$N$(H$_2$) and mass give the lower-limit values.  
Cloud positions are defined as the peak intensity position in 
galactic and equatorial coordinates (B1950).  The cloud mass is 
derived by summing up the $N$(H$_2$) of all the observed points 
where $W(^{12}$CO) is greater than 1.2 K km s$^{-1}$ (= 3 $\sigma$).  
The total mass of the 7 $^{12}$CO clouds is then calculated to be 
1630 $M_{\odot}$.  The Average $N$(H$_2$) is also estimated 
within the area where $W({\rm ^{12}CO}) \geq 1.2$ K km s$^{-1}$ 
for each cloud.  Peak $N$(H$_2$), $T_{\rm r}^*$, $\Delta V$, 
and $V_{\rm LSR}$ are measured at the most intense position of 
each cloud and $T_{\rm r}^*$, $\Delta V$, and $V_{\rm LSR}$ are 
estimated by fitting the spectral date to a single gaussian profile.  
The total mass of the whole observed area, where $W({\rm ^{12}CO}) 
\geq 1.2$ K km s$^{-1}$, is 2140 $M_{\odot}$.

\subsection{Distribution of $^{13}$CO clouds and C$^{18}$O cores}

Fig. 3 represents a comparison among the distributions of 
integrated intensity in $^{12}$CO, $^{13}$CO, and C$^{18}$O in L204 
complex.  The $J = $1--0 $^{13}$CO and C$^{18}$O spectra generally 
trace relatively denser regions of $\sim$ 10$^3$ cm$^{-3}$ and 
$\sim$ 10$^4$ cm$^{-3}$, respectively.  The $^{13}$CO emission is 
distributed as 3 distinct clouds which coincide with the previously 
identified clouds, S, T, and U by Nozawa et al.\ (1991), respectively.  
They are localized in the filamentary ridge of $^{12}$CO where 
$W({\rm ^{12}CO}) \geq 8$ K km s$^{-1}$.  The 3 $^{13}$CO clouds 
have masses of 500, 75, and 220 $M_{\odot}$, respectively.  
Tachihara et al.\ (2000) identified 8 cores in this region in C$^{18}$O 
(Fig. 3c).  These C$^{18}$O cores have clumpy shapes in general.  
Cloud S has an arc-like structure and the C$^{18}$O cores are aligned 
on the outer edge of the arc.  Cloud U has a ``bullet"-like structure and 
the C$^{18}$O core is located at the ``head".  The C$^{18}$O cores tend 
to be distributed relatively closely to $\zeta$ Oph in the $^{12}$CO and 
$^{13}$CO clouds.  Only one protostellar-like $IRAS$ point source, 
16442$-$0930, exists in this region and it appears to be associated 
with the core u2.  This suggests that star formation is taking place 
in the core.  The densest regions in these clouds have $N$(H$_2$) of 
$\sim 1.1 \times 10^{22}$ cm$^{-2}$ (peak position of core s5) as 
measured by C$^{18}$O.

\subsection{Velocity fields of the clouds}

The $^{12}$CO clouds show a velocity change from $\sim -3.0$ km 
s$^{-1}$ to $\sim 7.8$ km s$^{-1}$ and the velocity channel maps are 
shown in Figs. 4.  Each map shows integrated intensity of $^{12}$CO 
with a velocity width of 0.6 km s$^{-1}$.  Together with Fig. 5, it is 
seen that the 2 complexes have different velocity components.  The 
L156 complex has a more negative velocity ranging from $-3$ km 
s$^{-1}$ to 1.8 km s$^{-1}$, while L204 complex a more positive one 
from 1.2 km s$^{-1}$ to 7.8 km s$^{-1}$.  On the other hand, each 
complex has its own internal velocity structures.  

First, the L156 complex has the smallest velocity of $-2.0$ km 
s$^{-1}$ at the middle of the filament, $l \simeq 4^{\circ}\hspace
{-4.5pt}.\hspace{.5pt}5, b \simeq 22^{\circ}\hspace{-4.5pt}.\hspace
{.5pt}5 $, that is $\sim 2^{\circ}$ south from $\zeta$ Oph.  The 
velocity has a gradual change to 1 km s$^{-1}$ from this position 
of the apex in the velocity field to the both ends of the filament.  

Second, the L204 complex has a complicated velocity structure whose 
peak velocity ranges from 1.2 km s$^{-1}$ to 7.8 km s$^{-1}$ as a whole 
and each cloud has its unique velocity structure.  Cloud 2 has the 
largest velocity of $\sim$ 5 km s$^{-1}$ at the most western position 
(upper right in the figures) and the peak position changes to the east 
with the velocity decreasing down to $\sim$ 2.4 km s$^{-1}$.  Cloud 3 
has the largest velocity of $\sim$ 5.5 km s$^{-1}$ at the position where 
the C$^{18}$O cores s4 and s5 exist, and the intense regions change to 
$\sim$ 1.5 km s$^{-1}$ toward both sides along the filament.  
Cloud 5 has a velocity structure similar to cloud 2 and has the largest 
velocity of $\sim$ 5.4 km s$^{-1}$ at the position of the C$^{18}$O 
core, u1.  A general trend can be seen in the L204 
complex that the densest part of the cloud has the largest velocity 
in each cloud and the smaller velocity components spread toward 
the less denser regions away from $\zeta$ Oph.  These features show 
strong indication of the dynamical interaction between molecular 
clouds and $\zeta$ Oph as discussed in section 4.

\section{Discussion}

\subsection{Physical Interaction between the clouds and $\zeta$ Oph}

As shown by the morphology and velocity structure of the 
molecular clouds, the clouds seem to be interacting with $\zeta$ Oph 
and/or its surrounding H\,{\footnotesize II} region.  The clouds, cloud 
cores, and star formation are expected to be affected by $\zeta$ Oph.  
It is known by the previous studies that star formation is inactive 
around $\zeta$ Oph, although sufficiently dense and massive clouds 
and cloud cores exist (Nozawa et al.\ 1991; Tachihara et al.\ 2000).  
Magnetic fields are suggested to play an important role for cloud 
dynamics (McCutcheon et al.\ 1986, Heiles 1988) and the turbulent 
energy input from $\zeta$ Oph to the molecular clouds should 
be taken into account as well in considering star formation.  

In order to investigate the density and temperature distributions 
of a cloud qualitatively, we shall compare the distributions of 
$^{12}$CO, $^{13}$CO, C$^{18}$O, and 100 $\mu$m far-infrared 
emission.  Fig. 6 shows the intensity distribution of $IRAS$ 100 
$\mu$m emission overlaid with the contour of $^{12}$CO clouds.  We 
investigate the intensity distributions of the $^{12}$CO, $^{13}$CO, C$^
{18}$O, and 100 $\mu$m along 3 strip lines of A-B, C-D, and E-F 
illustrated by the blue lines in Fig. 6, as show in Figs. 7a-c.  The 3 kinds 
of CO distributions all show the steeper gradients toward $\zeta$ Oph.  
$^{13}$CO and C$^{18}$O are distributed locally close to $\zeta$ Oph and 
$^{12}$CO spreads to the opposite sides.  On the other hand, the 100 
$\mu$m peaks at the closer edge of the clouds to $\zeta$ Oph.  These 
features can be explained by the compression of the clouds by $\zeta$ 
Oph or by the surrounding H\,{\footnotesize II} region, denser cores are 
formed on the front side, and lower density gas may be accelerated to 
the opposite side toward the east.  The front sides of the clouds facing 
$\zeta$ Oph are illuminated by the UV light from $\zeta$ Oph, and the 
dust temperature may increase.  Thus the 100 $\mu$m emission is 
significantly enhanced on the front sides, suggesting that the gas and 
dust in this region may be compressed, accelerated and heated by 
$\zeta$ Oph.  

The 100 $\mu$m emission is relatively weaker toward the L156 
complex than toward the L204 complex as well as the CO intensity.  
The former lies on the H\,{\footnotesize II} region nearly across 
the center while the latter at the boundary.  The shadowing of the 
H$\alpha$ emission clearly shows that clouds are located in front of 
the H\,{\footnotesize II} region (Sivan 1974).  Because the L204 
complex has larger radial velocity and $N$(H$_2$) than L156 complex 
has, the following geometrical and kinematic structures are suggested; 
i.e., the 2 complexes are sheet-like molecular clouds sticking on the 
expanding H\,{\footnotesize II} region being pushed by $\zeta$ Oph.  
The L156 and L204 complexes are face-on and almost edge-on to us, 
respectively, and the projection effect makes the radial velocity and 
$N$(H$_2$) larger for the L204 complex.  If it is the case, the smallest 
velocity position on the L156 complex is the nearest to us.  Though it is 
actually $\sim 2^{\circ}$ south from the $\zeta$ Oph as mentioned 
above, it is explained by the rapid proper motion of $\zeta$ Oph.  
About $2.5 \times 10^5$ yr ago, $\zeta$ Oph existed by $\sim 30'$ 
closer to the position of the smallest velocity in L156 complex ($l 
\sim 4^{\circ}\hspace{-4.5pt}.\hspace{.5pt}5, b \sim 22^{\circ} 
\hspace{-4.5pt}.\hspace{.5pt}5$).  
The existence of some T Tauri stars (Nozawa et al.\ 1991) shows 
that the molecular clouds are preexistent and have formed stars 
prior to the passage of $\zeta$ Oph, and now we see the results of 
interaction between $\zeta$ Oph and the preexistent molecular clouds.  

As mentioned in Sec. 3.4, in L204 complex the lower density 
gas has smaller radial velocity and spread over far from $\zeta$ 
Oph.  This can be interpreted as molecular gas being pushed by some 
effects due to $\zeta$ Oph; gas at the near side of the cloud to $\zeta$ 
Oph is compressed and denser cores have been formed.  Some 
fraction of the gas has been tone off from the cloud.  The denser 
and more massive cloud cores can hardly be moved while the 
low-density gas can easily be accelerated.  It is likely that the L204 
complex had simpler velocity structures and that the gas and dust 
were not affected dynamically prior to the passage of $\zeta$ Oph.  
Since $\zeta$ Oph approached to the clouds, the complicated 
velocity fields may have resulted.  Around the inter-cloud region 
between the 2 complexes, faint string-like features in 100 $\mu$m 
running nearly perpendicular to the filamentary complexes can be 
seen, although no CO emission is detected perhaps due to low column 
density there.  These strings are also prominent in the east of the 
L204 complex as traced by the strip lines A-B, C-D, and E-F, and they 
are nearly parallel to the projected magnetic fields (McCutcheon et 
al. 1986).  This suggests that some of the gas and dust are tone 
off from the clouds and flowing to the downstream along the 
magnetic fields.  

We also note that the $^{12}$CO emission is significantly weak 
toward $\zeta$ Oph.  This may be because a large fraction of CO 
molecules are dissociated by UV light.  In the inter-cloud 
regions and cloud boundaries facing $\zeta$ Oph, there must be 
photo-dissociation regions (PDRs) and CO molecules cannot be 
detected from such regions.  This dissociation can also make the 
steep intensity gradients by the UV light penetrating into the 
cloud from the outside.  $N$(H$_2$) and the cloud mass 
traced by CO represent the quantities only for the dense part of 
the molecular cloud where UV flux is effectively shielded by 
interstellar dust.  In this sense, the obtained values of $N$(H$_2$) 
and mass should be regarded as the lower bounds as mentiond in 
Sec. 3.2.  Observations of other tracers like C\,{\footnotesize I} and 
C\,{\footnotesize II} emission with high spatial resolutions are 
required to estimate the gas density in PDR more accurately.

\subsection{Energetics of the cloud motion and the origin of the 
kinematic energy input}

To understand the physical interaction between $\zeta$ Oph and the 
molecular clouds quantitatively, kinetic energy and momentum of the 
clouds are investigated.  First, we assume that the gas component with 
the smallest velocity of cloud 1 and those with the largest velocity of 
clouds 2-7 are at rest and the gas with the other velocity component is 
flown away from the remaining gas due to the effects of $\zeta$ Oph.  
The momentum, $P$, and the kinetic energy, $E$, are estimated for 
each cloud as follows;
\begin{eqnarray}
P = \sum_{V=V_1}^{V_2} M_{V} |V-V_{1}| \ ,\\
E = \frac{1}{2} \sum_{V=V_1}^{V_2} M_{V} (V-V_{1})^2\ ,
\end{eqnarray}
where $V$ is the radial velocity, and $M_V$ is the mass contained in 
the channel at $V$.  Taking the $V_1$ and $V_2$ as the rest velocity 
and the velocity of the most highly accelerated component of each 
cloud, respectively, $P$ and $E$ are obtained as shown in Table 2.  
The total $P$ and $E$ of all the clouds amount to 2330 $M_{\odot}$ 
km s$^{-1}$ and $5.4 \times 10^{46}$ erg, respectively.  As the 
dynamical effect for the compression and acceleration of the gas 
due to $\zeta$ Oph, the stellar wind and UV radiation will be 
considered in the following.  

First, the energy and the momentum input by the stellar wind (SW) 
will be discussed.  If we accept the physical parameters of O9.5V 
type $\zeta$ Oph that the SW has $dm/dt \simeq 10^{-7} M_{\odot}$ 
yr$^{-1}$ and $V_{0} \simeq 1200$ km s$^{-1}$ (Morton 1975), 
where $dm/dt$ and $V_0$ are the mass-loss ratio and the escape 
velocity, respectively, the total energy and momentum of the SW for 
a period of $\tau$ yr are obtained as $E_{\rm SW} = \tau \times 
(1/2) (dm/dt) V_{0}^{2} = \tau \times 1.5 \times 10^{42}$ [erg] and 
$P_{\rm SW} = \tau \times (dm/dt) V_{0} = \tau \times 1.2 \times 
10^{-4}$ [$M_{\odot}$ km s$^{-1}$], respectively.  We assume here 
that clouds are of prolate shapes with the axial lengths listed in 
Table 1.  The area exposed to the wind is then expressed as $\pi ab$, 
where $a$ and $b$ are major and minor axes of the cloud, respectively.  
The energy and momentum input to the cloud is expressed as $E_{\rm 
SW} \times (\pi ab)/(4\pi d^{2})$ and $P_{\rm SW} \times (\pi ab)/
(4\pi d^{2})$, respectively where $d$ is the distance between $\zeta$ 
Oph and the cloud.  Thus, the expected time scales for energy and 
momenmum input ($\tau_{\rm E-SW}$ and $\tau_{\rm P-SW}$, 
respectively) are calculated independently for each cloud as listed 
in Table 2.

Next, we investigate the photo-dissociation effect by the UV 
flux of $\zeta$ Oph.  When a molecular cloud is illuminated by UV 
light, molecular gas will be evaporated away from the cloud with 
supersonic velocity, injecting momentum into the cloud.  
The cloud is then pushed backward as so-called the ``rocket effect" 
(e.g., Oort \& Spitzer 1954).  The kinetic energy input by the evaporated 
gas is roughly calculated as follows; the number of Lyman photons 
per each second radiated by $\zeta$ Oph is given as $1.2 \times 
10^{48}$ s$^{-1}$ (Panagia 1973).  When we assume that all the clouds 
are on the Str\"omgren sphere at $\sim$ 11 pc away from $\zeta$ 
Oph, the Lyman photon flux on the cloud surface, $J_{\rm L}$, is 
estimated to be $8.3 \times 10^7$ cm$^{-2}$ s$^{-1}$.  The particle 
density of ionized gas, $n_{\rm i}$, and $J_{\rm L}$ have a relation of 
\begin{equation}
J_{\rm L} = 0.01 \alpha n_{\rm i}^{2} r_{\rm i} \frac{m_{\rm i}^{2}+1}
{m_{\rm i}^{2}-1}\ ,
\end{equation}
where $m_{\rm i}$ is the Mach number of the streaming gas, 
$r_{\rm i}$ the spherical cloud radius, and $\alpha$ the recombination 
coefficient (Kahn 1969).  Here we take $m_{\rm i} = 2$, $\alpha = 2 
\times 10^{-13}$ (Reipurth 1983) and $r_{\rm i}^{2} = ab$, $n_{\rm i}$ 
can be calculated for each cloud.  The mass loss rate from a cloud 
hemisphere illuminated by UV is expressed as 
\begin{equation}
\frac{dM}{dt} = \pi ab \mu_{\rm i} n_{\rm i} v\ ,
\end{equation}
where $\mu_{\rm i}$ and $v$ are particle mass ($1.4 m_{\rm H}$) and 
escape velocity, respectively.  If we take $v = 20$ km s$^{-1}$ 
according to Oort \& Spitzer (1954), the mass loss rates of the clouds 
are estimated as listed in Table 2.  By using this, the energy input 
by photo dissociation can be expressed as $E_{\rm PD} = 1/2 (dM/dt) 
v^{2} \tau_{\rm E-PD}$ where $\tau_{\rm E-PD}$ is the time scale of 
photo dissociation.  Also for the momentum, $P_{\rm PD} = (dM/dt) 
v \tau_{\rm P-PD}$.  Assuming the kinetic energy and momentum of 
the clouds are made only by this rocket effect, we estimate 
$\tau_{\rm E-PD}$ and $\tau_{\rm P-PD}$ as listed in Table 2.  

The above two estimations are given as lower limits for the time 
scales because we assume that the momentum- and energy-transfer 
coefficients are 100\%, the tangential velocity of the flowing gas is 
neglected, and the UV shielding effect by dust grains is also 
neglected.  Nonetheless, these analyses tell us some information 
about the gas motion.  For some of the clouds, the energy input by SW 
needs a time scale, $\tau_{\rm E-SW}$, of only a few $\times 10^5$ 
yr or less, that is compareble with the crossing time of $\zeta$ Oph, 
$\tau_*$, through the clouds (see Fig. 5).  However, the time scales 
required for the momentum input range around $10^8$ yr and are 
significantly larger than $\tau_*$.  
Thus, we conclude that the stellar wind can hardly drive the high 
velocity gas in relatively short timescale of $\tau_*$.  An $U$-shaped 
bow shock around $\zeta$ Oph was found in [O\,{\footnotesize III}] 
(5010 \AA) and 60 $\mu$m bands (Gull \& Sofia 1979; Van Buren 
\& McCray 1988).  
This seems to be a partial shocked distorted bubble which is driven 
by SW.  The size of the bubble is expected to be smaller than a few 
pc, which is significantly less than the distance to the clouds, 
since the expansion time scale is only a few $\times 10^5$ yr 
(Weaver et al.\ 1977).  This also supports the result, because SW 
cannot affect the clouds penetrating through the bow shocks.  
On the other hand, the rocket effect by UV photo dissociation is 
more likely the cause of the peculiar velocity structures in the cloud.  
The required time scales for both the kinetic energies and momenta 
are less than $\tau_*$ and seem to be reasonable.  Here we 
conclude that the rocket effect may be one of the possible 
mechanism of the gas acceleration.

\subsection{Possible origin of the turbulence in molecular cloud}

Generally speaking, interstellar molecular gas is highly turbulent as 
inferred from the significantly larger molecular linewidth than the 
thermal one.  It is also a general trend that denser gas has a smaller 
linewidth, roughly speaking, $\sim$ 10 km s$^{-1}$ for 
H\,{\footnotesize I} gas, $\sim$ 2 km s$^{-1}$ for $^{12}$CO, $\sim 1$ 
km s$^{-1}$ for $^{13}$CO, and $\sim 0.5$ km s$^{-1}$ for C$^{18}$O.  
Recent studies revealed that the turbulent motion of gas may play 
an important role in cloud dynamics and star formation (Dobashi 
et al.\ 1996; Yonekura et al.\ 1997; Kawamura et al.\ 1998; 
Tachihara et al.\ 2000).  
Among all, Tachihara et al.\ (2000) suggest that the turbulence 
decay may lead to further contraction of cloud cores.  Since 
most of the cores in Ophiuchus North region are starless, there 
must be no disturbance from YSO formed in the cores.  As 
shown in detail by Nakano (1998), it is difficult to excite the 
turbulence in the cores from outside by magnetic field lines.  
Thus, the turbulence may be monotonically dissipated in time through 
the dynamical evolution of a core.  If this is the case, the amount of 
turbulent energy of the original diffuse gas cloud and the 
dissipation rate of the turbulence are essential factors in the core 
evolution.  As shown in the previous section, the photo 
dissociation by UV light is one of the probable causes of the 
turbulence.  Since $\zeta$ Oph is a run-away star crossing Ophiuchus 
region for about a few millions years, it may have disturbed the 
interstellar gas and put a significant amount of turbulent energy into 
all the clouds preexisting in the region during its travel.  There 
are a few tens of OB stars in the Sco OB2 association and the 
contribution of the rest of these OB stars should not be neglected.  
The field strength of the total UV radiation is estimated to be an order 
of magnitude larger than the typical value for the Galactic plane 
(Nozawa et al.\ 1991).  This may explain why the star formation 
in the Ophiuchus North region is very inactive.  
De Geus (1992) and Preibish \& Zinnecker (1999) present a 
scenario that a massive star in Upper Sco exploded as a supernova 
about 1.5 Myr ago.  A shock wave dispersed the clouds in the vicinity 
of the explosion and has compressed the $\rho$ Oph cloud core, 
inducing the extremely active star formation in the $\rho$ Oph cloud 
core.  We can see an expanding H\,{\footnotesize I} shell centered at 
$l \sim 345^{\circ}$ and $b \sim 25^{\circ}$, at whose boundary the 
$\rho$ Oph cloud exists (de Geus 1992).  The shock wave has not, 
however, reached the Ophiuchus North region (Tachihara et al.\ 1996).  
When we trace the position of $\zeta$ Oph back to 1.5 Myr ago, its 
former position shows good agreement with the center of the shell 
(de Geus 1992), and this also support the scenario that the previous 
companion of $\zeta$ Oph has exploded.

\section{Summary}

A summary of this paper is as follows:\\
1. $^{12}$CO ($J = $1--0) observations around the region of $\zeta$ Oph 
of 47 deg$^2$ have revealed two major filamentary cloud complexes 
(L156 and L204 complex) lying on the near-side of an H\,{\footnotesize 
II} region S27.  \\
2. These cloud complexes are divided into 7 clouds whose masses range 
from 70 $M_{\odot}$ to 520 $M_{\odot}$ and their total mass is 
1630 $M_{\odot}$.\\
3. The denser parts of the cloud which are traced by $^{13}$CO and 
C$^{18}$O emission are located in the filaments facing $\zeta$ Oph 
and the cloud complexes have peculiar velocity structures.  
These imply the physical interaction of the clouds and the 
H\,{\footnotesize II} region that the molecular gas is pushed and 
flown away from $\zeta$ Oph.  \\
4. We investigate the momentum and the kinetic energy for 
each molecular cloud.  The cloud momentum and kinetic energy 
range 60-800 $M_{\odot}$ km s$^{-1}$ and (0.9-21) $\times 
10^{45}$ erg, respectively.\\
5. As the origin of the streaming motion, the stellar wind from 
$\zeta$ Oph and UV photo dissociation effect are considered.  
The estimated time scales required for producing the momenta 
and kinetic energies of the clouds show that the stellar wind is 
hardly the cause of the gas acceleration while UV photo dissociation 
can be.  \\
6. The cloud turbulent motion is suggested to be an important 
factor for the cloud evolution by previous studies.  Our results 
indicate that the photo dissociation by a strong UV field may put 
the turbulent energy into the clouds and result in low star 
formation activity in this region if the contribution from other 
members of the Sco OB2 association is taken into account.

\par
\vspace{1pc} \par
We would like to thank Ralph Neuh\"auser, Yoshinori Yonekura 
and Nobuyuki Yamaguchi for very helpful comments on the manuscript.  
We greatly appreciate the hospitalities of all staff members of the Las 
Campanas Observatory of the Carnegie Institution of Washington.  
The NANTEN project is based on the mutual agreements between 
Nagoya University and Carnegie Institution of Washington.  
We also acknowledge that this project can be realized by the 
contribution from many Japanese public donators and companies.  
This work was financially supported in part by Grant-in-Aid for 
International Scientific Research from the Ministry of Education, 
Science, and Culture of Japan (No.10044076).
Two of the authors (YF and AM) acknowledge financial support from 
the scientist exchange program under bilateral agreement between 
JSPS (Japan Society for the Promotion of Science) and CONICYT (the 
Chilean National Commission for Scientific and Technological 
Research).

\clearpage
\section*{References}

\re
Bohlin, R.C. 1975, ApJ, 200, 402
\re
de Geus, E. J. 1992, A\&A, 262, 258
\re
de Geus, E.J., de Zeeuw, P.T., Lub, J. 1989 A\&A, 216, 44
\re
Dobashi, K., Bernard, J.P., Fukui, Y 1996, ApJ, 466, 282
\re
Draine, B.T. 1986 ApJ, 310, 408
\re
Gull, T.R., Sofia, S. 1979, ApJ, 230, 782
\re
Heiles, C. 1988, ApJ, 324, 321
\re
Hunter, S.D. et al.\ 1997, ApJ, 481, 205
\re
Kahn, F.D. 1969, Physica 41, 172
\re
Kawamura, A., Onishi, T., Yonekura, Y., Mizuno, A., Dobashi, 
K., Ogawa, H., Fukui, Y. 1998, ApJS, 117, 387
\re
Kopp, M., Gerin, M., Roueff, E., Le Bourlot, J. 1996, A\&A, 
305, 558
\re
Langer, W.D., Glassgold, A.E., Wilson, R.W. 1987, ApJ, 
322, 450
\re
Lesh, J.R. 1968, ApJS, 17, 371
\re
Lynds, B.T. 1962, ApJS, 7, 1
\re
Liszt, H.S. 1997 A\&A, 322, 962
\re
McCutcheon, W. H., Vrba, F. J., Dickman, R. L., Clemens, 
D. P. 1986, ApJ, 309, 619
\re
Morgan, W.W., Str\"omgren, B., Johnson, H.M. 1955, ApJ, 
121, 611
\re
Morton, D.C. 1975, ApJ, 197, 85
\re
Nakano, T. 1998, ApJ, 494, 587
\re
Nozawa, S., Mizuno, A., Teshima, Y., Ogawa, H., Fukui, Y. 
1991, ApJS, 77, 647
\re
Oort, J.H., Spitzer, L.Jr. 1955 ApJ, 121, 6
\re
Panagia, N. 1973 AJ, 78, 929
\re
Perryman, M.A.C. et al.\ 1997 A\&A, 323, L49
\re
Preibish, T. , Zinnecker, H. 1999 AJ, 117, 2381
\re
Reipurth B. 1983, A\&A, 117, 183
\re
Sivan, J.P. 1974, A\&AS, 16, 163
\re
Tachihara, K., Dobashi, K., Mizuno, A., Ogawa, H., Fukui, Y. 1996 
PASJ, 48, 489
\re
Tachihara, K., Mizuno, A., Fukui, Y. 2000 ApJ, 528, 817
\re
Van Buren, D., McCray, R. 1988, ApJ, 329, L93
\re
Weaver, R., McCray, R., Castor, J., Shapiro, P., Moore, R. 1977, 
ApJ, 218, 377
\re
Yonekura, Y., Dobashi, K., Mizuno, A., Ogawa, H., Fukui, Y. 
1997, ApJS, 110, 21

\clearpage
\centerline{Figure captions}
\bigskip

\begin{fv}{1}
{7cm}
{Integrated intensity map of $^{12}$CO ($J = $1--0) emission line in 
the Galactic coordinate.  Contours are drawn from 1.2 K km s$^{-1}$ 
with 2.4 K km s$^{-1}$ step.  Dotted lines show the observed area 
in $^{12}$CO.  Broken lines are the grids in the equatorial coordinate 
(B1950).  Plus mark denote the position of $\zeta$ Oph.}
\end{fv}

\begin{fv}{2}
{7cm}
{Identification of the molecular clouds.  Gray image is 
the integrated intensity of $^{12}$CO and the contours show the cloud 
boundaries.  The solid lines divide the connected molecular 
clouds.  Incompletely observed clouds are excluded.  Dotted 
lines show the observed area in $^{12}$CO.  }
\end{fv}

\begin{fv}{3}
{7cm}
{Close up images of L204 complex in $^{12}$CO (upper), 
$^{13}$CO (middle), and C$^{18}$O (lower).  The lowest contour levels 
and contour intervals are 1.2 K km s$^{-1}$ and 2.4 K km s$^{-1}$ for $^{12}$CO, 
1.6 K km s$^{-1}$ and 1.6 K km s$^{-1}$ for $^{13}$CO, and 0.18 K km s$^{-1}$ and 
0.18 K km s$^{-1}$ for C$^{18}$O, respectively.  The observed areas in 
$^{13}$CO and C$^{18}$O are shown by the broken lines.  $^{13}$CO clouds 
identified by Nozawa et al.\ (1991) and C$^{18}$O cores by Tachiahra et 
al. (1999) are denoted.  }
\end{fv}

\begin{fv}{4}
{7cm}
{Pseudo color images of channel maps in $^{12}$CO.  Each 
map is integrated over $\Delta V$ = 0.6 km s$^{-1}$.  The integrated velocity 
ranges are shown in each map.  Star in each map shows the 
position of $\zeta$ Oph.}
\end{fv}

\begin{fv}{5}
{7cm}
{Pseudo color image of peak velocity of $^{12}$CO emission.  
Cross mark and arrow show the position and proper motion of $\zeta$ 
Oph, respectively.  }
\end{fv}

\begin{fv}{6}
{7cm}
{Intensity map of $IRAS$ 100 $\mu$m.  Contours of $^{12}$CO 
cloud boundaries are overlaid.  Red circle show the extent of the 
H\,{\footnotesize II} region of S27.  Three blue lines denote the strips where 
intensity distributions are investigated (see text and Figs. 7).}
\end{fv}

\begin{fv}{7}
{7cm}
{Intensity distribution along the strip lines of A-B (a), 
C-D (b), and E-F (c).  Intensity of $^{12}$CO, $^{13}$CO, C$^{18}$O and 100 
$\mu$m are illustrated by solid, broken, dotted, and dash-dotted lines, 
respectively.  The intensity scales are shown in the left side ($^{12}$CO) 
and right side ($^{13}$CO and C$^{18}$O), while 100 $\mu$m is arbitrary 
scaled.}
\end{fv}

\end{document}